# Dragon-kings: mechanisms, statistical methods and empirical evidence

Didier Sornette [a,*], Guy Ouillon [b]

[a] *D-MTEC, and Department of Earth Sciences, ETH Zürich Kreuzplatz 5, CH-8032 Zürich, Switzerland (e-mail: dsornette@ethz.ch)*

[b] *Lithophyse, 4 rue de l'Ancien Sénat, 06300 Nice, France (e-mail: lithophyse@neuf.fr)*

**Abstract**
This introductory article presents the special Discussion and Debate volume "From black swans to dragon-kings, is there life beyond power laws?" We summarize and put in perspective the contributions into three main themes: (i) mechanisms for dragon-kings, (ii) detection of dragon-kings and statistical tests and (iii) empirical evidence in a large variety of natural and social systems. Overall, we are pleased to witness significant advances both in the introduction and clarification of underlying mechanisms and in the development of novel efficient tests that demonstrate clear evidence for the presence of dragon-kings in many systems. However, this positive view should be balanced by the fact that this remains a very delicate and difficult field, if only due to the scarcity of data as well as the extraordinary important implications with respect to hazard assessment, risk control and predictability.

## 1. Introduction

Many phenomena in the physical, natural, economic and social sciences can be characterized by power-law statistics, for which there are many different mechanisms (Mitzenmacher, 2003; Newman, 2005; Sornette, 2006). Power-law (i.e. self-similar) distributions of the sizes of events suggest that mechanisms of nucleation and growth dynamics remain the same over the whole spectrum of relevant spatial and temporal scales. According to this paradigm, small, large and extreme events belong to the same population, the same distribution, and reflect the same underlying mechanism(s). Major catastrophes are just events that started small and did not stop growing to develop into large or extreme sizes.

Following that reasoning, a majority of the scientific community considers those events as unpredictable, in the sense that the final size of a future event cannot be forecasted in advance. This is for instance the view espoused by P. Bak (1996) in the conceptual framework called "Self-organized criticality" introduced in Bak et al. (1987). This concept has become popular since Nassim Taleb's book "The Black Swan" (2007). This view is particularly pessimistic, and even alarming if true, as it casts a strong doubt on possibilities of precise hazard prediction, with all the societal consequences. In our mind, it is even dangerous as it promotes an attitude of irresponsibility: in a world where catastrophes (in human-controlled activities, for instance) are pure surprises, no one can be responsible. In contrast, the concept of dragon-kings, if their occurrences can be diagnosed ex-ante, brings back responsibility and accountability.

Are extreme events really due to the same mechanisms? Sornette (2009) argues that the power law paradigm may miss an important population of events that he refers to as "dragon-kings". Dragon-kings are defined as extreme events that do not belong to the same population as the other events, in a precise quantitative and mechanistic sense that will be developed below. The hypothesis articulated in (Sornette, 2009) is that dragon-kings appear as a result of amplifying mechanisms that are not necessary fully active for the rest of the population. This gives rise to specific properties and signatures that may be unique characteristics of dragon-kings.

For instance, in hydrodynamic turbulence, dragon-kings correspond to coherent structures or solitons developing in an otherwise incoherent noise background (L'vov et al., 2001). Dragon-kings could be related in some cases to critical transitions of the underlying dynamics. In such cases, extreme events would be somewhat predictable from the observation of the spatio-temporal dynamics of preceding small events.

There is growing evidence that some of the mechanisms that have been documented to be at the origin of power law distributions may, under slight changes of conditions and constraints, lead to log-normal distributions (Saichev et al., 2009) or to distributions with two regimes, such as dragon-kings, due to a transition from a kind of self-organized critical regime to a synchronized phase under increasing coupling strengths between constitutive elements (Sornette et al., 1994, 1995; Osorio et al., 2010).

The term 'dragon' stresses the idea that the mythological "animal" that a dragon embodies is endowed with mystical supernatural powers above those of the rest of the animal kingdom. The term 'king' (Laherrère and Sornette, 1998), refers to the fact that kings in certain countries are much more wealthy that the wealthiest inhabitants of their countries in a precise sense: while the wealth distribution over the whole population excluding the king obeys the well-known power law Pareto distribution, the king (including his family) is an "outlier", with a wealth many times larger than predicted by the extrapolation of the Pareto distribution. This "king" regime may result from cumulative historical developments, i.e., specific geo-political-cultural mechanisms that have enhanced the special king status in the wealth distribution (and also in other associated attributes that also contributed to the accumulation of wealth by the royal family).

---

*Corresponding author.
  E-mail address: sornette@ethz.ch

The dragon-king hypothesis has first been alluded to with the introduction of the term 'king' by Laherrère and Sornette (1998). It was then developed by Johansen and Sornette (1998a; 2001), who discovered dragon-kings in the distribution of financial drawdowns, proposed as a pertinent measure of risks (at that time, they referred to these dragon-kings as 'outliers'; this term is now abandoned to distinguish the dragon-kings, considered to be important parts of the dynamics, from the notion of statistical 'outliers', which usually refers to data that must be discarded because anomalous and surmised to be artifacts). Chapter 3 of (Sornette, 2003) provides a general review on the evidence for and importance of dragon-king, with an emphasis on their implication for financial market risks. A synthesis of the concept and proposed applications to seven different systems can be found in (Sornette, 2009; Satinover and Sornette, 2010).

The present special volume contains a series of articles that examine three classes of questions underlying the concept of dragon-kings:
(i) What are their mechanisms?
(ii) How to detect them? With what methods and/or statistical tests?
(iii) What is the empirical evidence for dragon-kings in various natural and social systems?

A fourth question on the predictability of dragon-kings is mostly left aside here. We briefly touch this subject at the end of this introductory article and refer to Sornette (2002; 2003; 2009) for progresses in this direction.

The systems, in which articles in the present special volume investigate the possible existence of dragon-kings, include: geosciences (earthquakes, volcanic eruptions, landslides, floods), economics (financial drawdowns, distribution of wealth), social sciences (distribution of city sizes), medicine (neuronal avalanches and epileptic seizures), material failure (acoustic emissions), hydrodynamics (turbulence, rain events, hurricanes, storms, snow avalanches) and environmental sciences (evolution and extinction of species, forest fires…).

The paper is organized into the five following sections: mechanisms and dynamics of dragon-kings (Section 2), detection of dragon-kings and associated statistical tests (Section 3), and empirical evidence of dragon-kings occurrence in a large variety of natural and social systems (Section 4). Section 5 discusses the potential implications of the dragon-kings concept in the predictability of major events. Section 6 calls for a new research and operational implementation of dragon-king simulators to endow decision makers with the tools to address the daunting challenges of the future crises.

## 2. Mechanisms for dragon-kings

### 2.1 Synthetic dragon-kings in random walks with long-range dependence

Werner et al. (2011) construct synthetic examples where dragon-kings can be detected in the auto-correlation function of complex time series of the random walk type. To make the problem relevant to complex systems that often exhibit long-range temporal dependence, they use the hierarchical Weierstrass - Mandelbrot Continuous - time Random Walk (WM-CTRW) model. The first type of dragon-king corresponds to a sustained drift whose duration time is much longer than that of any other event. This first abnormal event thus results from a mechanism that is similar in spirit to the abnormally large deterministic growth of Golosovsky and Solomon (2011) of stochastic proportional growth that is explained below. The second type of dragon-king takes the form of an abrupt shock whose amplitude velocity is much larger than those corresponding to any other event. Werner et al. quantify the durations of the abnormal drifts and the amplitudes of the abnormal shocks that lead to clearly differentiated signatures in the tail of the velocity autocorrelation function of the persistent random walks, or in other words, to clearly visible dragon-kings.

### 2.2 Abnormal deterministic grow leading to dragon-kings in a population following stochastic proportional growth.

Golosovsky and Solomon (2011) study the distribution of the citations up to July 2008 of 418,438 Physics papers published between 1980 and 1989. They show that the lognormal distribution model is adequate up to about 400 citations. This means that the number of Physics papers with citations less than about 400 citations have a frequency of occurrence well represented by the lognormal law. In contrast, Physics papers with citations between 400 and 1000-1500 are better described by a discrete power law model with exponent $\alpha \approx 2.15$ for the complementary cumulative distribution function. The lognormal and the discrete power models are found undistinguishable for papers with less than 400 citations, which is expected since the power law is asymptotically nested in a precise

---

*Corresponding author.
  *E-mail address*: sornette@ethz.ch

*Dragon-kings, black swans, and the prediction of crises* 3sense in the lognormal family (Malevergne et al., 2011a, 2011b). By comparison with the tests presented by Malevergne et al. (2011a, 2011b), the statistical technique of the 'uniformly most powerful unbiased' (UMPU) test applied to these two distributions would allow one to conclude that the power law model is overall preferred to the lognormal model.

Golosovsky and Solomon propose that the majority of papers have their citations evolve according to a stochastic proportional growth process. In the continuous limit, it takes the form

$$dS = \mu S\, dt + \sigma S\, dW, \qquad (1)$$

where the parameters $\mu$ and $\sigma$ quantify respectively the deterministic and stochastic components of the proportional growth. The term $dW$ is the increment of the standard Wiener process. Such process, together with some other natural ingredients such as birth and death, explains the power law distribution for a large part of the distribution (see Saichev et al., 2009 for a review of a large class of variants of proportional growth models).

However, from the point of view of this special volume, the most interesting contribution of (Golosovsky and Solomon, 2011) is the finding that the tail of the distribution for citations larger than 1500 cannot be explained by even the fat-tailed power law model. In other words, the empirical distribution of citations very strongly departs from the extrapolation of the power law model for numbers of citations larger than 1500. For the small set of papers that deviate from the power law model in the tail of the distribution, Golosovsky and Solomon find an abnormally large deterministic growth that make these articles different also in the characteristics of the dynamics of their citations. In the language of stochastic proportional growth processes (1), these special papers are characterized by a deterministic part $\mu dt$ much larger than the stochastic part $\sigma dW$. This is particularly interesting since the power law regime is in general emerging when exactly the reverse holds (Saichev et al., 2009; Malevergne et al., 2011a, 2011b). This anomalously strong deterministic component is diagnosed empirically by a strong temporal correlation of the growth dynamics of these papers. The abnormal growth and the evidence of a structure beyond the power law tail clearly qualifies these special Physics articles as forming a "dragon-king" class of its own.

### 2.3 Dragon-kings as "Bose-Einstein condensed droplets"

Yukalov and Sornette (2011) propose a novel theory explaining naturally the occurrence of dragon-kings in systems that tend to exhibit the Zipf law distribution. Consider the problem of determining the repartition of a population of N agents among C cities, according to their attraction and benefits that cities of different size may offer to their citizens. This problem is mapped onto the Bose-Einstein statistics of partitioning N particles among C energy levels. The cities are ranked by their size, where rank 1 corresponds to the large city, rank 2 the second largest, and so on. A control parameter $\beta$ is introduced to quantify the relative degree of attraction of cities according to their ranks. It is found that there is a critical value $\beta_c$ above which a finite fraction of the population condenses in the largest city of rank 1. As a consequence, exactly as for the Bose-Einstein condensate, the largest city does not follow, and stands beyond, the Zipf distribution of the other city sizes. Such a dragon-king coexists in a natural way with a power law distribution for the other cities, in the regime where the attraction to the largest members is larger than the critical threshold $\beta_c$. In this theory, there can be several increasing critical values $\beta_{c1}, \beta_{c2}, \ldots$ at which the largest cities successively condense away from the Zipf law.

### 2.4 Dragon-kings due to unsustainable transient herding in grand canonical minority games

Johnson and Tivnan (2011) study a simple model of a population of agents competing for a limited resource with bounded rationality. This is the so-called grand canonical minority game, previously studied extensively by Johnson and collaborators (Johnson et al., 2003). They report that extreme behaviors emerge due to an underlying mechanism that is fundamentally different from the mechanisms driving smaller changes. Moreover, the dynamics developing during these extreme events becomes more predictable than at other times, corresponding to transient bursts of dependence. These two properties are in line with Sornette (2009)'s description of the two main characteristics of dragon-kings.

The grand canonical minority game can be considered a stylized model of certain regimes of financial markets. Within this analogy, the dragon-kings are characterized by a quasi-deterministic inter-relationship between the number of active agents (proxy for the trading volume V) and the imbalance between the decisions being made (the excess demand D on the financial asset). Johnson and Tivnan (2011) classify different types of V and D trajectories, corresponding to the existence of a 'family' of dragon-kings.

The specific mechanism at the origin of the dragon-kings works as follows. In the situation where there are more agents than strategies available (the so-called "crowded regime"), it is likely that a given strategy can be held by many agents. When this strategy becomes successful, many agents will choose to activate it and hence buy or sell at the same time. This herding is mediated by the relative performance of the competing strategies with respect to the aggregate decision, which acts as a coordination mechanism. When such crowding occurs, this generates an extreme change, which is rather deterministic in nature. The change is transient because the strategy becomes rapidly a loser due to its being used by a growing majority.

**2.5 Positive feedback leads to finite-time singular shocks and dragon-kings**

The study of Yukalov et al. (2011) can be considered as a stylized conceptual extension of Johnson and Tivnan (2011), in that transient unsustainable herding is generalized into negative feedbacks of population on carrying capacity. Reduced form evolutionary equations for the evolution of populations of interacting agents or species emphasize the interplay between positive and negative feedback processes. The feedbacks are described via the impact of population on carrying capacity. This includes the self-influence of each kind of species on its own carrying capacity with delay. In the presence of destructive action of populations, whether on their own carrying capacity or on the carrying capacities of co-existing species, instabilities of the whole population may occur in the form of extreme events. These extreme events take the form of either finite-time booms followed by crashes or finite-time extinctions. Such behaviors of unsustainable super-exponential acceleration before collapse have been documented for many systems, from material rupture and earthquakes (Sammis and Sornette, 2002), to financial bubbles ending in crashes (Sornette, 2003; Jiang et al., 2010; Johansen and Sornette, 2010).

Akaev et al. (2011) apply mathematical models with such super-exponential growth to the dynamics of the Earth population and of commodities such as oil and gold. They show that the collapse of the oil price starting in July 2008 (a dragon-king following an accelerating bubble) could have been predicted up to one year in advance. It was actually predicted and reported ex-ante (Sornette et al., June 2008; 2009). Akaev et al. (2011) draw interesting if not terrifying scenarios concerning the future of human population and its impacts on economic development and social welfare in a comparative study of different world regions. One could summarize their analyses by saying that they expect dragon-kings to impact the dynamics of economies and nations.

**2.6 Dragon-kings as coherent collective events in the synchronized regime of systems of interacting threshold oscillators of relaxation**

In addition to their empirical analyses, Sachs et al. (2011) present numerical simulations of a system constituted of multiple slider blocks pulled over a surface at constant velocity, each block interacting with the surface through a static-dynamic friction law. Neighboring blocks are connected to each other by springs. Cascades of sliding events occur with a variety of sizes, where the size of a given event is quantified by the number of blocks that slip simultaneously. This type of model has a long history, starting with Burridge and Knopoff (1967), followed by the rediscovery by Carlson and Langer (1989), and the many papers in the 1990s that explored different variants and their properties. Increasing the stiffness of the system, the dynamics generates larger and larger events. For large stiffness, system-sized events occur, and stand clearly above the power-law distribution valid for smaller-size events. These system-wide events are clearly dragon-kings. In fact, the underlying mechanism is well understood and has been described previously by Schmittbuhl et al. (1993; 1996), Sornette et al. (1994), Gil and Sornette (1996) and Osorio et al. (2010) among others. In a nutshell, the block-spring system can be seen as a network of coupled threshold oscillators of relaxation. When the coupling strength (here quantified by the spring stiffness) is sufficiently large, the oscillators tend to synchronize into system-wide events. Sornette et al. (1994) and Osorio et al. (2010) have presented a general phase diagram in the plane (heterogeneity, interaction strength) that provides a powerful classification of when the synchronized regime occurs. In this regime, dragon-kings occur as system-wide events representing the tendency of the system to globally synchronize.

Sachs et al. (2011) also report the well-known fact that forest-fire models also exhibit system-sized events that occur with a frequency much larger than extrapolated from the distribution of smaller fires. These extreme events have been understood as collective coherent wave-like and spiral-like structures developing at large scales (Pruessner and Jensen, 2002; Bonachela and Munoz, 2009; Hergarten and Krenn, 2011). In this sense, they are as different from the rest of incoherent fires as are the coherent soliton-like velocity fluctuations found in shell models of turbulence different from the incoherent noisy velocity fluctuations forming the background (L'vov et al., 2001).

De Arcangelis (2011) explores the dynamics of neuron firing avalanches using numerical simulations. In agreement with reported experimental observations, she shows that such avalanches are power-law distributed in the case when inhibitory and excitatory mechanisms are in balance. However, when inhibitory mechanisms are depressed or when the neuron network features a very high connectivity level, very strong dragon-king events appear. More generally, under epileptic or hyper-excitable conditions, bi-modal event size distributions typically characterize an excess of large events that is inconsistent with the predictions from criticality. These large events that occur with very a high probability of occurrence compared with the extrapolation of the power law calibrated for the smaller events qualify as dragon-king avalanches. Experimentally, this regime is obtained by using pharmacological treatments that alter the balance between excitation and inhibition: picrotoxin reduces GABAergic inhibition, DNQX reduces fast glutamatergic synaptic transmission (AMPA receptors). The application of these drugs to cortex slices for a few hours is able to produce a dramatic change in the ongoing activity, which becomes supercritical (dragon-king regime) and subcritical, respectively. De Arcangelis notices that dragon-king avalanches are sensitive to slow NMDA rather than fast AMPA receptor antagonists during early development, which suggests that their presence is mainly controlled by fast GABAergic inhibition. Dragon-king avalanches appear in the cultures that realize a high level of burst synchronization. For instance, bicuculline increases burst synchronization leading to a large population of dragon-king avalanches. Dragon-king events may therefore affect the ability of the brain to give good performances in different tasks.

It should be stressed that the results of de Arcangelis concern only static properties measured within her particular model. A few experimental results on temporal correlations exist for the critical regime, which is associated with power law statistics, but not for the Dragon-Kings regime (see however the work of Plenz (2011) in this volume which is summarized below). This work echoes that of Osorio et al (2010) who studied the epileptic seizure distributions of 9 rats treated with low 3-MPA convulvant doses, corresponding to a weak coupling regime for which they observed power-law distributions. In contrast, the seizure energy distribution observed on 19 rats treated with maximally tolerable concentrations of 3-MPA, implying a strong coupling regime, resulted in a power-law behavior coextensive with characteristics large scale events qualifying as dragon-kings. This also led to the emergence of characteristics time scales in the dynamics. This reinforces the hypothesis that the coupling intensity and/or relative levels of excitation and inhibition control the emergence of dragon-kings in otherwise power law statistics.

Plenz (2011) argues that neuronal synchronization is essential to understand cortex dynamics in isolated preparations in vitro, in the awake animal in vivo, and in humans. Indeed, connections, i.e. synapses, between pyramidal neurons are relatively weak, when compared to the total input required to make a pyramidal neuron to produce an output. In addition, pyramidal neurons make only few connections with any given neuron, so that many neurons have to fire together to activate synapses at one downstream, i.e. postsynaptic, neuron.

In the presence of the tendency for neurons to synchronize their firing, neural networks tend to self-organize to produce spontaneous neuronal avalanches characterized by scale-free properties. This corresponds to a self-organization to a delicate critical point separating a sub-critical regime of small avalanches, when inhibitory substances dominate, from a super-critical regime of system-spanning avalanches, when excitatory transmission dominates. The properties of the self-organized scale-free neuronal avalanches include the power-law distributions of avalanche sizes and durations, the Omori-Utsu power law decay in number of avalanches aftershocks, and the productivity power law linking the size of an avalanche and the number of its triggered 'aftershocks'. These statistical properties are strongly reminiscent of those known for earthquake catalogs. As already mentioned, a similar correspondence has recently been documented for epileptic seizures (Osorio et al., 2010; Sornette and Osorio, 2010).

The self-organization to a delicate critical balance between excitatory and inhibitory synaptic transmission is remarkable. Indeed, even a small reduction in excitation changed the power law in avalanche sizes to an exponential size distribution, whereas a small reduction in inhibition led to bi-modal size distributions. The latter regime is characterized by a high degree of synchronization between neurons, leading to dragon-king avalanches. A possible mechanism for the convergence of the neuronal dynamics to a critical point was proposed by Sornette (1992) (see also Fraysse, Sornette and Sornette (1993) for proposed experimental implementations). The key idea is the existence of a non-linear feedback of the order parameter (average level of neuronal activity) on the control parameter (the balance between excitatory and inhibitory strengths), the amplitude of this feedback being tuned by the spatial correlation length. According to Sornette (1992), the self-organized nature of the criticality stems from the fact that the limit of an infinite (or system large) correlation length is attracting the non-linear feedback dynamics.

At the point of critical balance between excitatory and inhibitory synaptic transmission, cortical networks seem to optimally respond to inputs and to maximize their number of internal states. The dynamics of avalanches allows cortical networks to transmit and process information in an optimal way. The scale-invariant, balanced dynamics also allows for the emergence of a specific subset of avalanches, the coherence potentials. Coherence potentials form when the synchronization of a local neuronal group reaches a threshold, in which case the dynamics of the local group spawns replicas at distant network sites. Plenz (2011) suggests that the distinct, non-linear emergence of coherence potentials constitutes avalanche 'dragon-kings' of near perfect coherence. The functional importance of these coherence potentials stems from the proposal that they might constitute elements for a universal computing machine, analogously to gliders known to emerge in balanced cellular automata.

## 2.7 Delayed feedback loops for control of unstable systems

Delayed feedbacks to control instabilities often give rise to power law distribution of the amplitude of the control step. An example is stick balancing at the fingertip (Cabrera and Milton, 2011). Coexisting with a power law distribution of the motion amplitudes, Cabrera and Milton report the existence of maximum characteristic event sizes associated with the onset of fall, quite similarly to financial crashes (Johansen and Sornette, 2001; Sornette, 2003). Like the financial crashes, these maximum sized events may qualify as dragon-kings as they are likely to be associated with a change of the control system when the scale of the fluctuations in the controlled variable increases. Cabrera and Milton stress that the importance of the dragon-king hypothesis is that it shifts attention to the identification of the generating mechanisms for the fall, which marks the collapse of the stick and therefore the end of the dynamics. This is similar to the evidence that the incipient rupture event ending a progressive damage process within a material is a dragon-king, ending the life of the system (Sornette, 2009). The evidence for dragon-kings is further supported by the observation of a strong characteristic peak in the tail of the distribution of time interval between successive corrective movements (Cabrera and Milton, 2011). A possible model involves taking into account the interplay between feedback loops due to the mechano-receptor and visuo-motor control loops, which have different detection thresholds and different time delays acting onto a damped unstable oscillator.

## 3. Statistical methods to detect and qualify (or reject) the presence of dragon-kings

### 3.1 General remarks

The examples discussed in (Sornette, 2009; Satinover and Sornette, 2010) suggest several complementary methods for the identification of dragon-kings, which do not apply equally well to all systems. This identification can proceed by:
(1) Testing a reference model for the distribution of event sizes, and showing that the few extremes cannot be explained (fitted) by the distribution of the rest of the smaller population, using the theory of hypothesis testing and *p*-value; this approach has been used to diagnose dragon-kings in the distribution of financial drawdowns (Johansen and Sornette, 1998a; 2001);
(2) Identifying breakdown in the scaling of the distribution of event sizes at different coarse-graining resolution levels in the extreme tail, while the distributions collapse nicely for the rest of the distribution; this approach has been used to diagnose dragon-kings in the distribution of velocity increments in a shell model of turbulence (L'vov et al., 2001);
(3) Using normalization of event sizes with time-varying estimations of characteristic scales; this method has been used to diagnose dragon-kings in avalanche size distributions defined on a model of random directed polymers (Jögi and Sornette, 1998; Satinover and Sornette, 2010).

These three approaches are just three examples among many others that can be developed or already exist. In pure temporal systems, such as financial time series, it is sometimes necessary to take care of short time correlations instead of using a constant sampling rate. In spatio-temporal systems, the identification of dragon-kings also requires spatial segregations and space-time multiscale analysis. There is thus no universal method to diagnose Dragon-Kings, and other types of signals may necessitate still unexplored strategies.

We suggest that the exploration of the concept of dragon-kings in different systems can be organized by increasing complexity of the systems under investigations, starting with purely temporal signals, then considering purely 2D or 3D signals, to finally discuss problems with space-time-energy coupling.

The following subsections summarize the novel methods for the detection of dragon-kings that are introduced by different articles contributing to this special volume.

### 3.2 Pareto and tapered Pareto distributions as benchmarks

Schoenberg and Patel (2011) review the relative merits of (i) the Pareto distribution (or power-law distribution), used to describe environmental phenomena such as the sizes of earthquakes or wildfires, and (ii) the exponentially tapered Pareto distribution. For instance, Peters, Christensen and Neelin (2011) find that the later provides a good fit to the distribution of rainfalls taken indiscriminately as a whole. The motivation for the tapered Pareto distribution often stems from finite size effects (Cardy, 1988; 1992). In the case of earthquakes, a tail exponent $\alpha<1$ for the power law distribution of their energies would imply that the mean energy per earthquake is theoretically infinite or, in other words, the energy release rate is non-stationary, growing as $t^{(1/\alpha)-1}$. Clearly, such non-stationarity cannot be sustained over geological time scales, as this would imply the existence of earthquakes of magnitude corresponding to ruptures spanning of the whole Earth and more. This is obviously a physical impossibility: as a consequence, the power law distribution of earthquake energies has to be tapered and, in the end, bounded with a maximum upper limit (Pisarenko et al., 2008; 2010; 2011). The nature of the corresponding distribution remains a subject of intense investigations. Focusing on the pure Pareto and tapered Pareto distributions, Schoenberg and Patel (2011) review various studies showing that, even with rather large datasets, it is often quite difficult to distinguish them, because they differ markedly only in the extreme upper tail where few observations are recorded.

---
*Corresponding author.
 *E-mail address*: dsornette@ethz.ch

### 3.3 Tests for dragon-kings using confidence intervals on the empirical cumulative distribution function

Janczura and Weron (2011) propose a simple test to diagnose possible deviations in the tail of arbitrary distributions. The test is based on the fact that, in a sample of *n* events, the number of observations that exhibit a size smaller than some value *x* is binomially distributed with parameters *n* and $r=F(x)$, where $F(x)$ is the true cumulative distribution function (cdf) of the event sizes. The test is further simplified in its numerical implementation by noting that the central limit theorem allows for a much faster evaluation of the confidence intervals of the empirical cumulative distribution function at any desired point *x*.

There are two variations of the test. One consists in calibrating the whole empirical cdf with the target theoretical function constituting the null hypothesis, which can be a power law or a stretched exponential function (also known as Weibull), for instance. The second variation, which is preferred by Janczura and Weron, consists in fitting the theoretical function on the empirical cdf for the 10%-1% largest event sizes. In the former case, tail event sizes that fall outside the confidence intervals qualify as outliers with respect to the bulk of the distribution. In the latter case, tail event sizes that fall outside the confidence intervals qualify as outliers with respect to the 10%-1% tail of the distribution.

By using the latter approach of fitting the theoretical function on the empirical cdf for the 10%-1% largest event sizes, Janczura and Weron (2011) find no evidence of outliers for insurance claims associated with catastrophic losses from 1990-2004. In particular, the three largest events (Hurricane Andrew, the Northridge earthquake and the World Trade Center event of 11 September 2001) are seen as just tail events of a standard power law distribution. Similarly, none of the largest financial drawdowns of the NASDAQ index for the period Feb. 5, 1971-May 31, 2000 are qualified as outliers by this method, in contradiction with the conclusions of an earliest study by Johansen and Sornette (2001). However, when Janczura and Weron (2011) use the first variation of the test, they confirm the results of Johansen and Sornette (2001) and find that most of the 10% largest events are detected as outliers, that is, as being distributed with a distribution different from that of the bulk of the distribution. Janczura and Weron (2011) actually demonstrate that the 10% largest drawdowns are not in the confidence interval constructed on the basis of the fit of the whole distribution. According to this result, if only the 10% largest events are considered in their test (the second variation), being a homogenous group, it is clear that there should be no outliers within them. The fact remains that they have proven that the whole empirical cdf of drawdowns is distributed according to two regimes, roughly separated such that the 90% smallest drawdowns obey one stretched exponential distribution and the 10% largest ones obey another fatter tailed stretched exponential distribution or some other heavier tailed distribution. This second regime expresses the existence of other generating mechanisms at work, which correspond to the dragon-kings according to the terminology of this special volume. Similarly, coming back to the distribution of insurance claims associated with catastrophic losses from 1990-2004, when using the first variation of the test, i.e., when fitting the whole dataset with a Pareto law, the two largest claims do qualify as outliers (dragon-kings).

Analyzing electricity spot prices both in Germany and Australia, the test of Janczura and Weron (2011) qualifies a quasi-absence of dragon-kings in the price distribution in Germany, while dragon-kings systematically occur in Australia even under the restrictive application of the second variation of the test.

### 3.4 The U-test and the DK-test

Pisarenko and Sornette (2011) take up the daunting challenge of identifying a unique dragon-king in the tail of power law distributions. For this, they construct two statistics, leading to two tests dubbed the U-test and the DK-test, aimed at identifying the existence of even a single anomalous event in the tail of the distribution of just a few tens of observations.

The U-test calculates the p-value of the calibration of the power-law null hypothesis in an arbitrary interval of the random variable in question, i.e. with truncations both from above and from below. The later truncation corresponds to asking if the largest event(s) do(es) really conform to the same distribution as the rest. This procedure identifies both the presence of dragon-kings (higher estimated occurrence frequency) and the existence of tapering or downward roll-off (depletion of the most extreme events compared with the extrapolation of the null power-law).

The DK-test is derived such that the p-value of its statistic is independent of the exponent characterizing the null hypothesis, which can use an exponential or power law distribution. The DK-statistics is defined as the ratio of the average size (for an exponential null hypothesis) or log-size (for a power law null hypothesis) of events in the supposed dragon-king tail over the average size (or log-size) of events in the rest of the distribution.

By studying the distributions of cities and agglomerations in a number of countries, Pisarenko and Sornette (2011) demonstrate that these two tests do indeed provide evidence for dragon-kings in the distribution of city sizes of Great Britain (London), and of the Russian Federation (Moscow and St-Petersburg). Paris is diagnosed as a dragon-king in the distribution of agglomeration sizes in France. True negatives are also reported, for instance the absence of dragon-kings in the distribution of cities in Germany.

We feel that the U-test and the DK-test provide a straightforward generic methodology to check that the tail of a given empirical distribution abides to the distribution of the rest of the events. These two tests complement other approaches, such as those reviewed by Schoenberg and Patel (2011), by introducing statistics that are sensitive to deviations over just one or a few events, a treat that was often considered intrinsically impossible.

## 4. Empirical evidence for dragon-kings

### 4.1 Dragon-king in rainfall statistics

Süveges and Davison (2011) present a detailed statistical study of the catastrophic storm event that occurred in Venezuela on 14-16 December 1999, following an unusually wet fortnight. The storm struck the northwestern coast of Venezuela, bringing daily rainfall totals of 120 mm, 410.4 mm and 290 mm on three successive days at Maiquetia airport. This can be compared with the largest previously recorded event of 140mm since 1961. Standard extremal models cannot account for this catastrophic rainfall due both to inaccurate tail modeling and to an inadequate treatment of clusters of rare events. Süveges and Davison (2011) show that a mixture of distributions is needed (a so-called Dirichlet mixture model) in order to approximate the statistical data set since 1961 including the event of 14-16 December 1999, in terms of a seasonally varying mixture of types of extreme rainfall clusters. The statistical analysis classifies all the rainfall events recorded since 1961 into three types. The first type represents relatively long rainy periods, with variable profiles and a heavy tail distribution of rainfall. The two other types are concentrated on a single day and are finite-tailed. The extreme event on 14-16 December 1999 can thus be considered as a genuine dragon-king, pertaining to the heavy tail regime. In contrast, the other large rainfall events are of a different nature, due to local convective processes. The dragon-king may result from persistent orographic winds blowing from a relatively warm ocean surface meeting the coast. The existence of well-defined distinct generating processes creates precipitation with quite different characteristics, even though all may contribute to local weather patterns. Süveges and Davison (2011)'s results thus support the hypothesis that there are extreme events, like dragon-kings, which may be created by generative mechanisms that are very different from those holding for the majority of the other events (Sornette, 2009).

### 4.2 Hurricanes as dragon-kings

Peters, Christensen and Neelin (2011) study the statistical distribution of rainfalls in many regions of the Earth and show that ordinary precipitation events exhibit a universal power law distribution in both size measures, (i) event-depth and (ii) event power, i.e., rain rate integrated spatially over an event. For the largest sizes, the distribution falls off more rapidly than a power law, which reflects the finite water capacity of the atmosphere.

If one takes the naive definition of dragon-kings to be restricted to phenomena that stand distinctly above the power law in the size distribution, then there is no evidence for dragon-kings in rain events. But, a decade ago, L'vov, Pomyalov and Procaccia (2001) showed that dragon-kings can be hidden. Such dragon-kings are revealed by alternative signatures, such as different scaling laws linking the distributions of sizes at separate scales. Motivated by this insight, and given the special role played by hurricanes in the dynamics of the atmosphere, it is natural to ask whether hurricanes could indeed play the role of extreme events of a different physical nature with distinct statistical signatures, and thus be genuine dragon-kings.

Peters et al. (2011) stress the special physics associated with hurricanes. These coherent structures can be considered as genuine heat engines that connect the atmospheric surface boundary layer (which is warmed and moistened by ocean surface water) to the cold tropopause, with strong surface winds enhancing the transfer of energy from the sea surface. In contrast with ordinary cloud clusters that tend to fall apart more rapidly, hurricanes can sustain themselves over many days or even weeks. This is the signature of the fundamentally different dynamics by which high values of water vapor are maintained over a long time within hurricanes.

Following the reasoning that a different physical mechanism may indeed lead to visible statistical signature, Peters et al. (2011) develop independent criteria to distinguish hurricanes in a worldwide database. They show that, beyond the power-law regime in the region of the distribution of event sizes

---

*Corresponding author.
 E-mail address: dsornette@ethz.ch

where a roll-off occurs at an approximate exponential rate, hurricanes extend the tail in a manner that greatly increases the probability of the largest event sizes. In other words, the contribution to the tail of the size distribution of events associated with hurricanes falls much less quickly than that associated with other events. Peters et al. (2011) thus conclude that it is reasonable to use the term dragon-kings for hurricanes to refer to the fact that they are extreme in this measure, different from other events in morphology and longevity, and generated by a distinct physical process associated with a more efficient connection between the warm sea surface to the cold troposphere.

### 4.3 Dragon-king in rupture events

Lei (2011) shows that, in creep experiments on rock samples, the size distribution of all recorded acoustic emissions obeys a power-law, whereas the event corresponding to the rupture of the full sample defines a dragon-king, with a size much larger than what could be predicted using an extrapolation of the Gutenberg-Richter power-law magnitude-frequency distribution for either foreshocks or aftershocks. Similar observations were done by Nechad et al. (2005). Lei (2011) identifies two mechanisms for dragon-kings: 1) finite-time singular increase of the event rate and of the moment release (Sammis and Sornette, 2002); and 2) hierarchical fracturing behavior resulting from hierarchical inhomogeneities in the sample (Johansen and Sornette, 1998b; Sornette, 1998). In the first mechanism, the dragon-king event can be interpreted as the correlated percolation of many small events. In the second mechanism, an event of extreme size is the result of fracture growth stepping from a lower hierarchy into a higher hierarchy: rupture beyond a certain threshold size corresponding to the critical nucleation zone size is hard to stop.

Amitrano (2011) presents a numerical model of rock samples subjected to constant strain rate conditions. The sample is divided into cells, each of them being able to fail according to a Mohr Coulomb criterion with a random cohesion but a common internal friction. For low internal friction, the deformation is ductile-like at the macroscopic scale and becomes more and more brittle for model materials with larger internal friction. The distribution of energies released by micro-ruptures behaves as a power law for small events with anomalously large events (dragon-kings) in the brittle regime.

Amitrano documents significant variations of the distribution of rupture energies in his model that he compares with empirical observations at various scales (laboratory tests, field experiments, landslides, mining induced seismicity, crustal earthquakes). A perfect power-law fits the whole range of rupture scales (excluding the incipient macro-rupture) of his numerical model for an internal friction of $\mu \approx 0.5$, in the range of values for crustal materials. Amitrano suggests that this could be the origin for the non-systematic observation of characteristic events in earthquake sizes.

Despite the macroscopic change of rheology with internal friction, Amitrano observes that the final sample failure displays typical signatures of a critical point (Sornette, 2006). The ductile-brittle transition can be considered as the progressive appearance of dragon-kings, i.e. extreme events of a different nature and/or statistics. Despite their unpredictability in size, considering they obey the critical point theory, these particular events remain predictable in time, at least if a temporal evolution of the control parameter is known, as they emerge from the divergence of the failure process.

It is worthwhile to review the history of what is meant by "critical point theory" (Alava et al., 2006) for material failure and to provide key references, because its existence underpins the fact that dragon-king events correspond to the incipient critical rupture ending a progressive damage phase. Critical point theory for rupture means that the material under stress is undergoing a progressive damage process, with micro-rupture events revealed as acoustic emissions whose energies are power-law distributed up to an upper cut-off that increases as the system gets closer and closer to the global incipient rupture. The dragon-king corresponds to the final global rupture run-away whose energy release is orders of magnitudes larger than the most extreme events during the damage phase.

The ancestor of the critical earthquake concept is probably (Vere-Jones, 1977) who used a branching model to illustrate the concept of rupture cascades. Allègre et al. (1982) proposed what amounts to be a percolation model of damage/rupture prior to an earthquake. But they phrased the model using real-space renormalization group theory and critical phenomena, emphasizing the multi-scale nature of the physics. Their real-space renormalization group approach to their effective percolation model recovers the previous work of Reynolds et al. (1977) and translates it into the language of seismology. Smalley et al. (1985) presented another similar real-space renormalization group theory of the stick-slip behavior of faults, as a model of cascades of earthquakes. Sornette and Sornette (1990) use the critical earthquake concept put forward by Allègre et al. (1982) to propose an empirically observable verification of the scaling rules of rupture. Voight (1988; 1989) is probably the first to introduce the idea of a time-to-failure analysis in the form of a second order nonlinear differential equation, which, for certain values of the parameters, leads to a solution of the form of time-to-failure equation (i.e., a power-law divergence of strain or some of its time derivatives). He did use it later for predicting volcanic eruptions. Sykes and Jaumé (1990) was probably the first paper reporting and quantifying with a specific law an acceleration of seismicity prior to large earthquakes. They used an exponential law to describe the acceleration and they did not introduce the concept of a critical earthquake. In contrast, Bufe and Varnes (1993) introduced a power-law relationship between a measure of deformation or damage at time t and the time distance $t_c-t$ to the occurrence of a global rupture at $t_c$. Specifically, they proposed that the cumulative Benioff strain $\varepsilon_B$ is proportional to $1/(t_c-t)^m$, where the exponent m is positive and $t_c$ is the time of the occurrence of the target earthquake. They provided fits to real time series obtained from seismic catalog to (post-)predict the occurrence of large target earthquakes. Their justification of the relation $\varepsilon_B \sim 1/(t_c-t)^m$ was a mechanical model of material damage. They did not speak of and did not introduce the concept of a critical earthquake. Sornette et al. (1992) and Sornette and Sammis (1995) reinterpreted the work of Bufe and Varnes (1993) as well as all previous ones and generalized them within a statistical physics framework to spell out explicitly the concept of a critical earthquake. They described how a large earthquake could be seen as a genuine critical point. Using the insight of critical points in rupture phenomena, Sornette and Sammis (1995) proposed to enrich the equation $\varepsilon_B \sim 1/(t_c-t)^m$ by considering complex exponents (ie log-periodic corrections to scaling). Extensions can be found in (Anifrani et al, 1995; Saleur et al., 1996; Johansen et al., 1996; Huang et al., 1998; Bowman et al., 1998; Jaume and Sykes, 1999; Johansen et al., 2000; Ouillon et al., 2000; Zöller and Hainzl, 2001; Sammis et al., 2004; Shebalin et al., 2006; Lei and Satoh, 2007; Keilis-Borok and Soloviev, 2010). See also Mignan (2011) for an exhaustive review of Accelerating Moment Release before large events and alternative, more reductionist interpretation schemes.

The critical earthquake concept is in fact associated with a parallel development in the statistical physics approach of material failure, called the critical rupture concept. Roots of this approach are found in the book of Herrmann and Roux (1990) on scaling and multiscaling of rupture in heterogeneous media. The time-dependent description of the critical rupture concept can be found in (Sornette and Vanneste, 1992; Vanneste and Sornette, 1992; Lamaignère et al., 1996; Andersen et al., 1997; Sornette and Andersen, 1998; Johansen and Sornette, 2000; Ide and Sornette, 2002).

**4.4 Extraordinary snow avalanches as dragon-kings**

Ancey (2011) presents a review of case studies of extreme snow avalanches that support the hypothesis that some extraordinary avalanches can be termed genuine outliers or dragon-kings, because they cannot be predicted from the extrapolation of distributions calibrated on past observations and/or using standard models. These extraordinary avalanches extend beyond the predictions of existing zoning procedures. While errors could have been made in mapping extreme avalanches or the knowledge of the physical processes may be insufficient, their detailed observations suggest to Ancey (2011) that some extreme avalanches have their own dynamics, which makes them physically and mechanistically different from other avalanches. Ancey defines a dragon-king as an avalanche whose features (volume, impact pressure, distance, and so on) are markedly different from other rare events.

Ancey suggests four main processes, which can sometimes work in combination, to explain the occurrence of dragon-king avalanches: (i) change in bed path roughness by filling gaps in the topography, (ii) trajectory switch, (iii) diversion by former deposits, and (iv) self-induced deforestation. Some avalanches may modify the roughness of the avalanche path by filling gaps in the topography (or by destructing forests). Such morphological changes, possibly coupled with the inertia of a large avalanche, may also change the trajectory of the snow mass and allow the occurrence of an extreme avalanche even in case of moderate to small snowfall. Ancey (2011) also points out that any quantitative work on avalanches is difficult as each event is a multivariate process whose full quantification is difficult to assess. He also notes that statistics must be made considering avalanches occurring within a full mountain range to get enough data, as avalanches propagating along the same path are not numerous enough. A pure power-law distribution of snow avalanche sizes may thus emerge even if some dragon-king events occur within single paths, in a way similar to seismicity or landslides analyses.

**4.5 Uncertain evidence of dragon-kings for earthquakes**

In their paper, Sachs et al. (2011) present a short review of empirical distributions of event sizes in a variety of systems, from earthquakes, volcanic eruptions, landslides, to wildfires and floods. Main and Naylor (2011) review the evidence for characteristic earthquakes (a notion equivalent to dragon-kings) in comparison with material failure in the laboratory and volcanic eruptions.

Sachs et al. report that the distributions of sizes of landslides and wildfires are well described by power-laws and the largest events do not seem to deviate from the power law (no abnormal behavior and no dragon-kings). This holds also for volcanic eruptions, notwithstanding the occurrence of very large past volcanic eruptions, such as Mount Tambora (Indonesia), 1815, V=150km$^3$ and Lake Toba (Indonesia), 73,000BP, V=2,800km$^3$ (where V is the volume of erupted material). These events are still on the extrapolation of the power law distribution calibrated on the crowd of smaller events. If anything, the distribution tends to bend downward, exhibiting a depletion of extreme volcanic eruptions as can be expected from the finite size of the Earth (this is called a finite-size effect in statistical physics and scaling analysis). In contrast, Main and Naylor (2011) note that some volcanic sequences show clear characteristic behavior with well-defined strain precursors allowing for rather precise advance warnings, and that nearly all laboratory tests on rock samples show an extreme event (a dragon-king) at the sample size, well beyond the population of acoustic emissions that largely indicate grain scale processes until very late in the cycle.

In contrast, Sachs et al. suggest that certain tectonic regions may exhibit dragon-kings. To show this, they consider a sub-catalog featuring all 1972-2009 events that occurred in a zone fitting with the location of aftershocks of the 2004 Parkfield event in central California. They point out that the extrapolation of the Gutenberg-Richter law to large magnitudes 'predicts' that the largest event should have a magnitude approximately equal to M=5.65, whereas the largest event that occurred on September 28, 2004 had M=5.95. This discrepancy is taken as the signature of a dragon-king, the so-called "characteristic" Parkfield event. This conclusion is highly controversial and deserves much more scrutiny than offered by Sachs et al. (2011). In order to understand the nature of the difficulties, let us briefly retrace its history and recent developments.

The characteristic earthquake hypothesis was proposed by Schwartz and Coppersmith (1984), and documented by Wesnousky (1994 and 1996). Because of its importance in seismic hazard evaluation, the hypothesis surely deserved careful study and rigorous testing. Recall that it was still underlying the fault segmentation used for the hazard calculations in the region at the time of the March 2011 Tohoku earthquake. This approach led to a dramatic under-estimation of the seismic hazard in the Tohoku region, so that an earthquake of M=9.0 was indeed thought to be essentially impossible in that region. However, before the Tohoku event, the characteristic earthquake hypothesis has been challenged on the basis of pure statistical arguments (Page, 2009).

Actually, the characteristic earthquake hypothesis has encountered many problems during its testing. Kagan and Jackson (1991) tested McCann et al.'s model (1979) by studying earthquakes that occurred after 1979. They found that large earthquakes were more frequent in those zones where McCann et al. (1979) had estimated low seismic potential. The characteristic model was considerably refined by Nishenko (1989, 1991) who applied it to enough zones in a prospective forecast that could be tested at a high confidence level. Kagan and Jackson (1995) found that earthquake data after 1989 did not support Nishenko's model. Rong et al. (2003) concurred. Jackson and Kagan (2006) discussed the characteristic model and its testing in more detail. The characteristic hypothesis was vigorously contested in the Nature debate (1999) on earthquake prediction. Most of seismologists seem either to ignore the test results or explain them by Nishenko's fault. Since it was shown that previous applications of characteristic earthquake hypothesis ended in a clear, unambiguous failure, and no new testable (falsifiable) models have been proposed since the mid 1990s, it seems fair to say that current applications of the characteristic earthquake hypothesis cannot be supported.

Main and Naylor (2011) conclude that the evidence for characteristic earthquakes is far from compelling, though their existence cannot be ruled out either. At present, there is no irrefutable evidence for dragon-kings in the sense of outliers from the power-law size distribution of earthquake event sizes. In contrast, studying the empirical distribution of earthquake energies in California from 1990 to 2010 (extract from the ANSS catalog), Amitrano (2011) finds that the magnitude distribution for events larger than M=6 displays a significant discrepancy compared to the power-law, so that these larger events could be considered as characteristic earthquakes.

Main and Naylor stress that, by their very nature of allowing large fluctuations, power law distributions imply very large variable sizes of the largest events in a given finite sample. Sometimes, the largest event drawn out of a power law distribution could happen to be 10 times larger or more than the second largest event, which would lead to naively conclude for the existence of a dragon-king (in the case of Sachs et al., the largest observed event energy is only twice the one of the largest 'predicted' one). This underlines the difficulties of identifying genuine dragon-king events. In the case of Sachs et al., the seismicity during the chosen time period (which fits with the assumed seismic cycle duration for this fault segment) may be dominated by aftershocks of the 2004 event. More precisely, it is possible to quantify the typical gap in size between the largest event and the second larger event drawn from a power law distribution. This is called Bäth's law in seismology. Using the self-excited Hawkes conditional Poisson model (known as the ETAS model in seismology, for epidemic-type aftershock sequence), Helmstetter and Sornette (2003) have shown that the magnitude gap of about 1.2 between a main shock and its largest aftershock can be accounted precisely by combining the power law distribution of earthquake energies with the scaling law for the dependence of fertility (number of aftershocks as a function of the magnitude of the main shock). Main and Naylor thus warn that great care must be taken in assigning significance to potentially illusory 'dragon-kings' identified by a magnitude gap or by comparing the observed frequencies with the trend extrapolated from the distribution at smaller magnitudes. The search itself for dragon-kings may introduce a hidden bias that must be taken into account in statistical tests of the significance of a large magnitude gap. Two powerful tests addressing this question are introduced by Pisarenko and Sornette (2011), as mentioned above. It anyway remains that the bottleneck for a definitive conclusion is the independent and objective determination of a fault, given that they are subparts of larger, self-similar networks, and consist in the aggregation of smaller-size sub-faults. New clustering techniques that reverse engineer fault networks from seismicity (Ouillon et al., 2008; Ouillon and Sornette, 2011) may provide the basis to treat the earthquake-fault chicken-egg problem consistently (Sornette, 1991).

**4.6 Uncertain evidence of dragon-kings for floods**
Using the example of the distribution of flood sizes, Sachs et al. (2011) illustrate that the definition of a dragon-king might be model dependent, but their conclusion has to be taken with caution as we now explain. They show a plot of the maximum yearly discharge Q as a function of the return period T for the Colorado River in the Grand Canyon, Arizona. The log Pearson type 3 distribution, used as a standard in US forecasts, is found to fit reasonably well the flood size frequency data recorded at the Lees Ferry gauging station from 1921 to 1962, except for the largest event in the series that clearly departs from the rest. To this data, Sachs et al. add two even larger paleo-floods, (A) the historic Colorado River flood of 1884 with discharge estimated at Q = 8,800 m$^3$/s and recurrence interval of T = 112 years and (B) an event that occurred 1200-1600 years BP with an estimated discharge Q = 14,000 m$^3$/s and recurrence interval T = 1400 years. Then, they suggest that a power law could fit these three events. Upon examination of the presented figure, it is difficult to uphold their claim that the power law could also account for a part of the distribution of the other smaller floods. In contradiction with their conclusion that there are no dragon-kings because a power law can account for the tail of the distribution, we interpret the data as evidence for two populations of floods, one approximately described by the thin-tailed log Pearson type 3 distribution and possibly a second (power law or other fat-tailed) distribution for the three largest floods. But even this conclusion is quite uncertain due to the addition of the events (A) and (B) that occurred over a time span of more than three millennia to a database of just 42 years. Indeed, consider the following. For a distribution (pdf) of floods $p(F) \sim 1/F^{1+\alpha}$, the largest flood over $n$ years scales typically as $F_{max}(n) \sim n^{1/\alpha}$. Consider a time period of $N \gg n$ years. Then, the largest flood over these N years is of the order of $F_{max}(N) \sim N^{1/\alpha} \gg F_{max}(n) \sim n^{1/\alpha}$. By this reasoning, it can be expected that floods (A) and (B) that are recorded over a time scale $N$ must be anomalously large compared with the set of floods recorded over the period $n \ll N$. These events are not genuine dragon-kings, only statistical artifact due to mixing of data sets of different durations, and extraction the largest events in the longest dataset and putting them in the shorter dataset. Perhaps the largest flood in the 42-year data set can qualify as a dragon-king but this needs to be ascertained by rigorous statistical tests that can apply to a single abnormal event, for instance using the method proposed by Pisarenko and Sornette (2011). In conclusion, the question of the existence of dragon-kings for floods remains open.

**4.7 Some systems with plausible dragon kings that are not studied in this special volume**

**4.7.1 Rogue waves**
Rogue waves, also called freak waves, begin with a deep trough followed by a wall of water reaching 25-30 meter high. They occur in deep water where a number of physical factors such as strong winds and fast currents converge, in the presence of different mechanisms that include interference effects and non-linear interactions and modulation resonance. We refer to Heller (2005) and in particular to Akhmediev and Pelinovsky (2010). This special volume reviews the evidence and possible mechanisms for rogue waves occurring in oceans, plasmas, nonlinear optical systems, and Bose-Einstein condensates. These studies suggest that rogue waves are indeed outliers compared with the rest of the fluctuations and result from specific initial conditions and/or particular amplifying mechanisms. Rogues waves are thus candidate dragon kings.

**4.7.2 Substorms in the ionosphere**
The global energy transport in the solar wind-magnetosphere-ionosphere system around the Earth is characterized by a large variability characterized by a power law dependence of the power spectrum of auroral indices and by power law distributions of the energy fluctuations of in situ magnetic field observations in the Earth's geotail. These power laws have been suggested as evidence for scale free self-organized criticality (SOC) in analogy with sandpile avalanche models (Bak et al., 1987). In contrast, the intensity of, and time interval between, substorms both have well defined probability distributions coexisting with characteristic scales for the most extreme events. In other words, two types of dissipation events are found: (i) those internal to the magnetosphere occurring at all activity levels with no intrinsic scale, and (ii) those associated with active times corresponding to global energy dissipation with a characteristic scale (Chapman et al., 1998; Lui et al., 2000). Simple cellular automata have been introduced to model these energy fluctuations as avalanches (Chapman, 2000). The power law probability distribution of energy discharges is suggested to be due to internal reorganization in agreement with the SOC paradigm, whereas system-wide discharges form a distinct group with a characteristic scale. This suggests the existence of distinct processes acting in the generation of the largest events in the dynamic geotail.

### 4.7.3 Human parturition

In the hours, days, weeks and even months before the actual childbirth occurs, the musculature of the uterus begins to momentarily tense and relax. As the coupling between muscle fibers increases, these muscle fibers tend to synchronize more and more, ending in the parturition itself, which corresponds to the crossing of the critical value at which a synchronization of all muscles of the uterus occurs, leading to the "travail" associated with regular periodic large scale contractions and the birth of the baby. The transition or bifurcation to the parturition phase can be therefore considered as providing an example of a dragon king embodied in the "travail" phase. In collaboration with obstetricians, Sornette et al. (1994; 1995) and Vauge et al. (2000) have developed the first steps of a human parturition model based on the concept of bifurcation, which views the true labor in childbirth as the post-critical phase associated with a DK in the amplitude and duration of Braxton-Hicks contractions. As the muscles spasms, called contractions, may or may not be the leading indicators of imminent labor and the resulting childbirth, monitoring these contractions to construct a predictive model may aid in differentiating between contractions that indicate the forthcoming childbirth and those known as false labor, a.k.a. Braxton-Hicks contractions. Furthermore, knowledge of the different types of contractions would help identify high-risk pregnancies and possibly predict an impending premature birth.

### 4.7.4 Large positive deviations for power law distributions

Marsili (2012) has recently remarked that large positive deviations for power law distributed variables occur by the fact that a finite fraction of the whole sample deviation turned out to be concentrated on a single variable. Recall that large deviations are random events such that their sample averages do not take the values prescribed by the law of large numbers (i.e. the expected value), but rather deviate considerably from it (Gnedenko, 1998; Sornette, 2006, chapter 3).

One can summarize this mechanism for such a dragon-king to occur as follows. Consider a random variable x, with a probability density function (pdf) given by $p(x) \sim 1/x^{1+\alpha}$ with a given exponent $\alpha$. Suppose that N realizations ($x_1, x_2, \ldots, x_N$) of the random variable are recorded and that the realized moment of order q, $\langle x^q \rangle_{empiric} = (1/N) [x_1^q + x_2^q + \ldots + x_N^q]$ with $q < \alpha$, is different from its unconditional theoretical expectation $\langle x^q \rangle_{theory}$. Two cases occur.

If $\langle x^q \rangle_{empiric}$ is smaller than $\langle x^q \rangle_{theory}$ by a difference of order 1 (and not of the standard order of magnitude $\sim 1/N^{1/2}$), the empirical pdf p(x) has to be truncated by an exponential taper. This example is illustrated in Chapter 3 of (Sornette, 2006) for the case of earthquake energies, which are distributed according to the Gutenberg-Richter distribution corresponding to $\alpha \approx 2/3$. Then, imposing the condition that the average energy (moment with q=1) released by the Earth has to be finite leads to the truncation of the Gutenberg-Richter distribution by an exponential taper (Schoenberg and Patel, 2011).

If $\langle x^q \rangle_{empiric}$ is larger than $\langle x^q \rangle_{theory}$, one can show that the Cramer function is zero, which means that the distribution is unchanged but there is one event, the largest one, which is singular and of sufficient amplitude to change just by itself the expected $\langle x^q \rangle_{theory}$ into the observed $\langle x^q \rangle_{empiric}$ (Marsili, 2012). This mechanism is reminiscent of the Bose-Einstein condensation mechanism proposed by Yukalov and Sornette (2011) in this volume. Note that, if the conditioning is imposed on the log-moment, then the exponent of the conditional distribution is decreased from its unconditional value $\alpha$, similarly to a standard "democratic" large deviation result, but here acting homogeneously across all the population, changing its structures by changing the exponent of the distribution. Note also that the large positive deviation dragon-king mechanism can be mapped onto a condensation mechanism occurring in the particles-in-box model with Hamiltonian equal to the logarithm of the sum of the number of particles in each box. Then, the density as a function of chemical potential shows a phase transition for a density larger than a critical value, for which all boxes except one have an average number equal to that for zero potential and one box has a finite fraction of the total number of particles, while the distribution of particles per box is a power law with exponent $\beta$, the inverse temperature (Bialas et al., 1999; Evans and Hanney, 2005; Evans et al., 2006). There are potentially many applications to this positive large deviation result for power laws, including the nature of large biological extinctions and the structure of financial crashes.

## 5. Predictability of extreme events

Dragon-kings events obviously belong to the subset of events that we would wish to forecast or predict. Forecasting itself is generally split into subcategories such as time-dependent and time-independent forecasts. Time-independent forecasts aims at providing the probability of occurrence of a given event in a finite time-interval, assuming that future activity does not depend on present or past activity. Time-dependent forecast aims at the same result using extrapolations of more or less probable scenarios to future times. Prediction aims at providing an accurate estimation of the size, location and time of occurrence of a future event, generally based on a careful analysis of precursory activity (see chapter 10 of Sornette, 2006; Dakos et al., 2008; Scheffer et al., 2009).

An outstanding challenge is to classify the phenomena discussed in this special volume according to their degree of predictability. If dragon-king events do not stem from a bottom-up process and are solely controlled by the size of the system, they may be distributed in time according to a homogeneous, Poisson-like process. In that case, they would be unpredictable but high quality (time-independent) forecasts are possible. This is for example the idea standing behind the characteristic earthquake concept as defined by classical seismology with the elastic rebound theory of Read (1910). If large events stem from amplification mechanisms at smaller hierarchical levels in the system, they might be preceded by a precursory activity allowing their prediction, or forecasted using time-dependent algorithms (such as the ETAS model (see Helmstetter and Sornette (2002) for a review and references therein). In that case, precursors may be qualified as 'active'. Note, however, that the tectonic elastic rebound theory has been recently revisited to include the possible complex signature of smaller-scale events (see Mignan, 2011, for a review). This shows that even if small events do not significantly interact to prepare collectively the much larger upcoming event, they can nevertheless be used to predict it as passive tracers of the oncoming singularity created by the large event, despite the system being possibly far from a critical state.

Another goal consists in the definition of the domain or sub-domain within which precursors should occur, if any. In the case of 1D problems (such as market rates developing as a function of time), the identification of precursors involves relatively straightforward metrics, such as the price dynamics itself and that of the volatility and transaction volumes (Sornette, 2003). In the case of rock or material samples brought to failure in the laboratory, we can still reduce the dynamics to a pure temporal one (as we are only interested in the time of global failure – this is also the case for volcanic eruptions). Things are not that easy in the case of space-time dynamics of very large systems when one is interested in the prediction of the local dynamics. In that case, one has to guess not only the spatial location of the future event, but also the size of the domain within which precursors to that event will occur. For example, in retrospective analyses, several authors used to look for earthquake precursors within domains that were much larger than the size of the 'predicted' event (see Bowman et al., 1998, and Mignan, 2011, for a review). However, as dragon-kings seem to be identifiable only at the scale of individual faults for the case of the distribution of earthquakes, this suggests that, perhaps, precursory patterns should occur more often at that local scale too (at least for seismic precursors). We thus meet again the necessity of defining unambiguously what a fault is. This problem of spatial domain definition is indeed generic in any time-space dynamic system and needs to be rigorously tackled in the future, using sophisticated pattern recognition techniques, for instance.

Making progress in the domain of prediction also requires to review successes and failures of prediction (and forecast) in case histories, and to re-interpret them in the light of what we learned about the different possible types of dragon-kings. The phase diagram proposed by Sornette (2009) is an interesting starting point, provided we deal with fields whose different regimes can be reducible to those mapped in this diagram. One important question is to characterize what is meant by coupling or interaction strength and what is heterogeneity. Different systems may also have different relevant parameters that are not reducible to just coupling strength versus heterogeneity. Other such phase diagrams may be necessary as their structure may change with the loading conditions and the influence of external factors and systems. For example, predicting large events in mechanical systems may necessitate to carefully scrutinize the space and time evolution of the 3D, second order strain tensor. The dimensionality of the phase diagram may increase accordingly.

In terms of prediction, in the scenario in which dragon-kings are associated with a phase transition or more generally a bifurcation or change of regime, we should not look especially at the largest 'instantaneous' events, but also at the largest correlated sequences of small or medium-sized events, or at any other major change in the dynamics of the system. In the case of earthquakes, for instance, this suggests to look at large or giant swarms, without any empirically clearly identifiable 'main' event, and not only at the potentially destructive 'big-one' (in the spirit of Johansen and Sornette, 1998a). This reasoning could

---

*Corresponding author.
  *E-mail address*: dsornette@ethz.ch

apply to many other fields. Sornette (2002; 2006), Dakos et al. (2008), Scheffer et al. (2009) and others have documented that the theory of bifurcations and phase transitions predicts the presence of early warning signals for the upcoming transition into the major regime shifts that can be called dragon-kings. These signals include

(i) a slowing down of the recovery from perturbations,
(ii) increasing or decreasing autocorrelations,
(i) increasing variance of endogenous fluctuations,
(ii) appearance of flickering and stochastic resonance, and other noise amplification effects (Harras et al., 2011),
(iii) increasing spatial coherence, and
(iv) singular behavior of metrics revealing positive feedbacks (Sammis and Sornette, 2002; Johansen and Sornette, 2010).

To the overarching question that motivated initially this special volume, "Is there life beyond power laws?", the cumulative knowledge provided by the articles that follow indeed confirms the existence of a lively and exciting domain of research, with many surprises and developments to be found in the near future. We thus wish a long and fruitful life to the investigations aimed at discovering, explaining and predicting the often crucial and revealing deviations from power law distributions that we have called "dragon-kings".

## 6. Control of extreme events

To end this review, it is warranted to ask how much the understanding obtained on dragon-kings could lead to operational utility. In this vein, we would like to push for a vigorous extension of the program introduced by Dorner et al. (1990) and Dorner (1997) to develop simulation platforms that incorporate detailed physical geological, meteorological, geological, architectural and economic data with all known (and to be tested) feedback loops. In the simulators developed by Dorner et al., decision makers are allowed to make decisions on allocated resources to develop projects and to mitigate risks according to different strategies. The simulations demonstrate the consequences of the decisions within a multi-period set-up. The perhaps not-so-surprising results were that humans experience failure more often than success when intervening in complex systems. The key point however is that the failures are not random but exhibit common patterns, including: lack or decreasing questioning of assumptions over time; insufficient prior analysis; failure to anticipate side effects; incorrect interpretation of the system's reaction (no immediate obvious negative effect wrongly interpreted as "all is well"); over-involvement in pet projects distracting from the main objectives; and so on.

These are eerily familiar, when transposed to the causes of the subprime crisis that started in 2007 with epicenter in the US, to the on-going sovereign crisis in Europe, to the Deep Horizon BP oil spill disaster in 2010, to the Challenger explosion in 1986, to the maiden flight in 1996 of the European Ariane 5 rocket, and so on. We live in a world where central banks are performing experiments in real time that are impacting billions of people, based on (so-called Dynamical Stochastic General Equilibrium) models that, until recently and after the crises, did not even incorporate a banking sector and could not even consider the possibility of systemic financial failures due to contagion. Not much has changed, though. We suggest that changes to this "primitive" approach are sorely needed in order to endow decision makers in financial, policy, engineering and environmental domains, as well as the public, students and anyone interested and responsible, with tools to learn and to practice at the level that airline pilots or surgeons already experience in their training, i.e., with "flight" or ``surgery'' simulators.

While Dorner and his collaborators have obtained important insightful results, we believe that their efforts have only scratched the surface of the potential offered by this type of simulation approach. With the specific angle on bifurcations and dragon-kings, we believe that extending Dorner et al.'s simulators on a systematic and operational level is perhaps the most pressing challenge of modern times. Indeed, with the responsibility of steering the planet towards sustainability at a juncture when, according to the World Wide Fund for Nature of the UNESCO, mankind is already consuming the equivalent of 1.5 planet Earths in *renewable* resources, we need to make choices that do not rely on our cognitive biases, behavioral quirks, political games and limited ability to learn from history. We need to develop science-based methods to help incorporate responsible and efficient decision making into our "DNA". We believe that this can only be achieved with bifurcation/dragon-kings simulators. Only by "living" through scenarios and experiencing them can we make progress. There is enormous evidence in the laboratory and in real life settings that veterans who have lived through bubbles and crashes, through environmental crises and so on, are much better at prevention and mitigation. But the cost is too large to learn from real life crises. We believe it is possible to develop simulators for decision makers to understand the complex dynamics of out-of-equilibrium systems whose behavior intrinsically includes changes of regimes, bifurcations, tipping points and their associated dragon-kings. The decision maker thus first needs to understand the dynamics of his system holistically, in a systemic way, which means that he needs to understand the existence of dragon-kings as one of the dynamical solutions of the evolution of his system. He needs to have a classification of the different regimes possible, a phase diagram in which he understands which control leads to the region of the dragon-kings and which do not. He needs to understand that bifurcations and changes of regime are a natural and expected part of natural and social systems. For most people, this understanding does not occur via studying fancy mathematical models of the type we have described above and in this volume but, instead, by experimenting as in real life, albeit with the protective luxury of the simulator and the efficiency of scaling space and time as needed. Only under this systemic structural understanding, can the decision maker interpret correctly the precursory signs of dragon-kings in real life and use them to correct and steer the system to sustainability.

---
*Corresponding author.
*E-mail address*: dsornette@ethz.ch